# Steady-State Simulation for Combined Transmission and Distribution Systems

Amritanshu Pandey[1], *Graduate Student Member,* Larry Pileggi[1], *Fellow, IEEE*

*Abstract*— The future electric grid will consist of significant penetration of renewable and distributed generation that is likely to create a homogenous transmission and distribution (T&D) system, requiring tools that can model and robustly simulate the combined T&D networks. Existing tools use disparate models and formulations for simulation of transmission versus distribution grids and solving for the steady-state solution of the combined T&D networks often lacks convergence robustness and scalability to large systems. In this paper, we show that modeling both the T&D grid elements in terms of currents and voltages using an equivalent circuit framework enables simulation of combined positive sequence networks of the transmission grids with three-phase networks of the distribution grids without loss of generality. We further demonstrate that we can ensure robust convergence for these resulting large-scale complex T&D systems when the circuit simulation methods are applied to them. Our results illustrate robust convergence of combined T&D networks using a direct Newton-Raphson solver on a single machine for smaller sized systems and using a parallel Gauss-Seidel-Newton solver on multiple machines for larger sized systems with greater than million nodes.

## I. INTRODUCTION

Future electric grid will strive for most efficient and economical use of grid resources while maintaining high reliability and safety. The grid today is changing with growing adoption of variable and intermittent sources of generation, especially wind and solar, in the power systems across the globe. These high levels of penetration of renewables will result in much narrower operational margin than what's available today, thereby significantly affecting the reliability of the grid. To ensure that the reliability is not affected, interdependencies between the transmission grid and distribution grid (wherein a significant fraction of solar is likely to be installed) will have to be clearly understood while enabling control based on knowledge of the operating state for both the transmission and distribution systems. This was apparent when a transmission system operator in PJM coordinated with the Sturgis, Michigan distribution grid to avoid a blackout by utilizing 6 MW of distributed generation back in 2013 [1]. To securely and reliably enable control actions such as this, the operators and planners of the grid will require new simulation capabilities that will navigate through the invisible boundaries that exist today between the transmission and distribution (T&D) grid analyses and solution methodologies while simulating high penetration of renewables and Distributed Energy Resources (DERs). The existing simulation frameworks for power system analyses are unable to capture these interdependencies between the T&D grids. No standard tool exists today that can jointly model the large interconnected T&D grids while ensuring robust convergence of the steady-state model for the same. This lack of simulation capability was highlighted in an ARPA-E workshop to identify paths to large-scale deployment of renewable energy resources, where one speaker noted that the "tools are not graceful in considering penetration levels at which much of the thermal fleet could get de-committed," and that "studies do not co-simulate impact of renewable injection into receiving AC systems" [2]. Another speaker noted that the tools for simulating increasingly coupled T&D systems "are not well integrated" [3]. Additionally, other industry experts have noted the lack of tools that can simulate the effect of high penetration of DERs in the system [4]-[5]. For e.g. author in [4] notes that the existing commercially available software tools may be inadequate to accommodate high penetration of DERs in the system.

Many frameworks exist for analysis of combined T&D networks including those for obtaining steady-state [6]-[10] and time-domain transient [11] and those applying optimization methods [12]-[13]. Amongst these, steady-state analysis (power flow methods) of combined T&D networks is the most consequential for planning and operation of the grid. However, the existing approaches for this analysis lack convergence robustness especially for large sized systems. This is primarily due to the use of disparate formulations and algorithms for modeling and analysis of transmission versus distribution networks [6]-[9]. The most common practice for steady-state analysis of combined T&D grid is to model the transmission network via a positive sequence model and the distribution network via a three-phase model, then to couple the two. This relies on a fundamental assumption that the three-phases of the transmission network are balanced at the point of interconnection (POI). In general, most of these methods tend to couple the transmission and distribution systems via an interface and then solve the two via disparate simulation tools [7]-[9] in a *co-simulation framework*. For instance, [7] models the transmission grid via PowerWorld and the distribution grid via GridLab-D. The co-simulation is then performed by running individual sub-circuits in their respective tools and then by exchanging variables via a communication port. Similar approaches are also used in [8]-[9],[14]. However, due to the use of disparate tools/methods for solving the individual T&D

This work was supported in part by the Defense Advanced Research Projects Agency (DARPA) under award no. FA8750-17-1-0059 for the RADICS program.
[1]Authors are with the Electrical and Computer Engineering Department, Carnegie Mellon University, Pittsburgh, PA 15213 USA, (e-mail: {amritanp, pileggi}@andrew.cmu.edu).

test cases, it is difficult to develop methods that are both generalized and able to guarantee convergence for both transmission and distribution systems.

As an alternative approach, *combined T&D framework* can be used where the aggregated T&D network is solved in the same framework/tool. This is demonstrated in master-slave approach [6], wherein the aggregated network is modeled and solved in a common distributed framework. In this method, the problem is split into a transmission power flow and several distribution power flow sub-problems that are then solved via different algorithms to capture the different features of T&D grids. However, the methodology has mostly been tested on small scale systems with no claims of robust convergence for the individual sub-systems. Another approach for combined T&D simulation is to model all the three phases of the transmission grids and then to couple the same with three-phase networks of the distribution grids [10]. This approach does not require a balanced operation assumption of the transmission grid, and thus allows for modeling of unbalanced conditions for transmission grid as well. However, the primary limitations of this approach are the general lack of three-phase data for the transmission network, and the lack of research toward ensuring robust convergence of three-phase transmission networks.

In this paper, we propose a novel framework for combined T&D simulation that uses equivalent circuit formulation to jointly model the T&D grid without loss of generality and ensure robust convergence for the steady-state simulation of the same via the use of developed circuit simulation methods. The primary contributions of the paper address the challenges in the existing frameworks for combined T&D simulations [14] and are as follows:

i) The proposed framework can solve for the steady-state of any complex combined T&D network robustly independent of the choice of initial conditions.
ii) The proposed framework is highly scalable and can solve extremely large sized systems including systems with greater than millions of nodes and is only limited by the availability of computation infrastructure.

To robustly solve the combined T&D system independent of the size or complexity of the network, we develop a direct Newton Raphson (NR) solver for smaller sized systems to run on a single core as well as a parallel Gauss-Seidel-Newton (GSN) solver for large-sized systems to run on distributed cores on a server or on cloud. In the results section, we solve for the steady-state solution of combined T&D networks with greater than million nodes.

## II. Equivalent Circuit Formulation

The equivalent circuit approach [15]-[17] for steady-state analysis of power systems has been applied separately to transmission and distribution power grid problems to address the scalability, convergence robustness, and modeling limitations of the respective existing formulations. This approach for generalized modeling of the power system in steady-state (i.e. power flow and three-phase power flow) represents both the T&D power grid elements in terms of equivalent circuit elements that can be coupled without modification to further perform combined T&D steady-state analysis. Importantly, our objective is a formulation that can represent any physics based model or measurement based semi-empirical model as a positive sequence or three-phase sub-circuit, as shown in [18], [19] and [20], and that these models can be combined hierarchically with other circuit abstractions to build larger combined T&D aggregated models. These aggregated circuits can be further abstracted by a graph $G$ with a set of system nodes $V$ and series elements $\mathcal{E}$. We start with a PQ load model to demonstrate formulation of the positive sequence and three-phase equivalent circuit for the same. Following the approach shown in this section and in [17], positive sequence or three-phase equivalent circuit of any power grid element can be developed.

### A. Positive Sequence Model for the PQ bus

In the equivalent circuit approach, the positive-sequence load (PQ) bus model is modeled via a complex current source [16] that is a function of complex voltages ($V_{RL}$ and $V_{IL}$, respectively). To enable the application of NR, this complex current source is split into real and imaginary current sources ($I_{RL}$ and $I_{IL}$, respectively). This is necessary due to the non-analyticity of complex conjugate functions [17]. The resulting equations for the positive sequence PQ model for power flow problem are:

$$I_{RL} = \frac{P_L V_{RL} + Q_L V_{IL}}{(V_{RG})^2 + (V_{IG})^2} \quad (1)$$

$$I_{IL} = \frac{P_L V_{IL} - Q_L V_{RL}}{(V_{RL})^2 + (V_{IL})^2} \quad (2)$$

When NR is applied to solve the corresponding nonlinear system of equations, the components of the system, which include the PQ buses, are linearized, solved, and updated in an iterative manner. Following the equivalent circuit approach in [17] the positive sequence model for a PQ bus used in power flow is shown in Fig. 1 for the (k+1)$^{th}$ iteration of NR. It is constructed by linearizing the set of equations (1)-(2) for the positive sequence parameters and then representing the resulting equations using fundamental circuit elements.

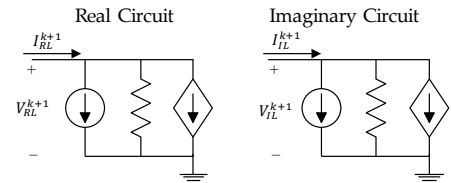

*Figure 1: Equivalent Circuit Model for PQ load model.*

### B. Three-Phase Model for the PQ bus

Similar to the positive sequence model for the PQ bus, the three-phase model with individual phases can be represented via a set of controlled current sources [17]:

$$I_{RL}^{\Omega} = \frac{P_L^{\Omega} V_{RL}^{\Omega} + Q_L^{\Omega} V_{IL}^{\Omega}}{(V_{RL}^{\Omega})^2 + (V_{IL}^{\Omega})^2} \quad (3)$$

$$I_{IL}^{\Omega} = \frac{P_{RL}^{\Omega} V_{RL}^{\Omega} - Q_L^{\Omega} V_{RL}^{\Omega}}{(V_{RL}^{\Omega})^2 + (V_{IL}^{\Omega})^2} \quad (4)$$

where, $\Omega$ represents a phase from the set of phases $\{a, b, c\}$.

To represent the three-phase PQ model current sources as an equivalent circuit for each NR iteration, we first linearize (3)-(4) to obtain a set of linearized current sources. Since the three-phase loads in the grid are usually connected in wye-connection or delta connection, the linearized circuits can be connected as such depending on the type of load connection. Fig. 2 represents the mapped real equivalent circuit for the three-phase PQ load model when connected in wye formation and in delta formation.

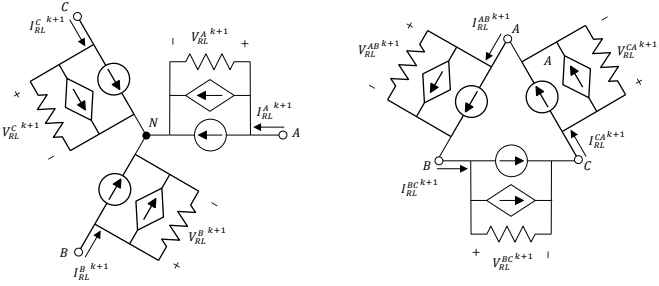

*Figure 2: Real circuit for a) wye connected ZIP Load Model (on left) b) delta (D) connected ZIP load model (on right).*

Following the approach shown above and in [17], the positive sequence or three-phase equivalent circuit of any power grid element can be constructed. In the following sections, we develop the methods used for robust convergence of these circuits, then introduce the new approach for combined T&D simulation.

## III. CIRCUIT SIMULATION METHODS

Circuit formalism and decades of research in circuit simulation field have demonstrated that circuit simulation methods can be applied for determining the DC state of large-scale, highly non-linear circuits using NR. These techniques have been shown to make NR robust and practical for circuit problems [21] consisting of billions of nodes. Most notable is the ability to guarantee convergence to the correct physical solution (i.e. global convergence [22]) and the capability of finding multiple operating points [23]. We have previously proposed analogous techniques for ensuring convergence to the correct physical solution for the power flow problem and three-phase power flow problem [17]. Here we extend these methods to be used with combined T&D steady-state analysis. Note that throughout the paper, the symbol superscript $\Omega$ in the mathematical expressions of the equivalent circuit models represents a phase from the set $\Omega_{set}$ of three phases $a$, $b$ and $c$ for the three-phase sub-circuit model and represents the *positive sequence* ($p$) component for the corresponding transmission sub-circuit model.

### A. General Methods
#### 1) Limiting Methods

A simple, yet effective, technique for improving convergence of NR is to limit the absolute value of the delta step at each iteration. For nonlinear power system problems, this corresponds to the real and imaginary positive-sequence and three-phase voltage vectors that are updated for each NR iteration in the combined T&D simulation. This is analogous to the voltage limiting technique used for diodes in circuit simulation, wherein the maximum allowable voltage step during NR is limited to twice the thermal voltage of the diode [24]. The mathematical implementation of voltage limiting for combined T&D simulation is as follows:

$$(V_C^\Omega)^{k+1} = \min_{V_C^{min}} \max_{V_C^{max}} \left( (V_C^\Omega)^k + \delta_S \min\left( \left|\Delta(V_C^\Omega)^k\right|, \Delta V_C^{max} \right) \right) \quad (5)$$

$$\min_{V_C^{min}} \max_{V_C^{max}} = \begin{cases} V_C^{max}, & \text{if } x > V_C^{max} \\ V_C^{min}, & \text{if } x < V_C^{min} \\ x, & \text{otherwise} \end{cases} \quad (6)$$

where $\delta_S = \text{sign}\left(\Delta(V_C^\Omega)^k\right)$ and $C \in \{R, I\}$ represents the placeholder for real and imaginary parts. We have demonstrated the efficacy of such limiting methods for large-scale power flow and three-phase power flow problems in [17].

### B. Homotopy Methods

Limiting methods may fail to ensure convergence for certain ill-conditioned and large combined T&D test systems when solved from an arbitrary set of initial guesses. To ensure convergence for these network models to the correct physical solutions, independent of the choice of initial conditions, we make use of homotopy methods. Here we discuss one such homotopy method called "Tx stepping" [25].

In Tx stepping method, the series elements in the system (i.e. set of transmission lines ($\mathcal{T}_X$) and transformers (xfmrs)) are first "virtually" shorted to solve the initial problem that has a trivial solution. Specifically, a large conductance ($\gg G_{il}$) and a large susceptance ($\gg B_{il}$) are added in parallel to each transmission line and transformer model in the system. In case of three-phase power flow, a large self-impedance $(\gg Y_{\Omega\Omega}^{il})$ is added in parallel to each phase of the transmission line and transformer model. Furthermore, the shunts in the system, are open-circuited by modifying the original shunt conductance and susceptance values. Importantly, the solution to this initial problem results in solution with high system voltages (magnitudes), as they are essentially driven by the slack bus(es) complex voltages and the PV bus voltage magnitudes due to the low voltage drops in the lines and transformers (as expected with virtually shorted systems). Similarly, the solution for the bus voltage angles will lie within an epsilon-small radius around the slack bus angle. Subsequently, like other continuation methods, the formulated system problem is then gradually relaxed to represent the original system by taking small increment steps of the homotopy factor ($\lambda \in [0,1]$) until convergence to the solution of the original problem is achieved. Mathematically, the line and transformer impedances during homotopy for the positive sequence model within the combined T&D network is expressed by:

$$\forall il \in \{\mathcal{T}_X, \text{xfmrs}\}: \hat{G}_{il} + j\hat{B}_{il} = (G_{il} + jB_{il})(1 + \lambda\gamma) \quad (7)$$

and for the three-phase model:

$$\begin{bmatrix} \hat{Y}_{aa}^{il} & \hat{Y}_{ab}^{il} & \hat{Y}_{ac}^{il} \\ \hat{Y}_{ba}^{il} & \hat{Y}_{bb}^{il} & \hat{Y}_{bc}^{il} \\ \hat{Y}_{ca}^{il} & \hat{Y}_{cb}^{il} & \hat{Y}_{cc}^{il} \end{bmatrix} = \begin{bmatrix} Y_{aa}^{il}(1+\gamma\lambda) & Y_{ab}^{il} & Y_{ac}^{il} \\ Y_{ba}^{il} & Y_{bb}^{il}(1+\gamma\lambda) & Y_{bc}^{il} \\ Y_{ca}^{il} & Y_{cb}^{il} & Y_{cc}^{il}(1+\gamma\lambda) \end{bmatrix} \quad (8)$$

where, $G_{il}$, $B_{il}$, and $Y_{\Omega\Omega}^{il}$ are the original system impedances and $\hat{G}_{il}$, $\hat{B}_{il}$, and $\hat{Y}_{\Omega\Omega}^{il}$ are the system impedances used while iterating from the trivial problem to the original problem. The parameter $\gamma$ is used as a scaling factor for the conductances and susceptances. If the homotopy factor ($\lambda$) takes the value one, the system has a trivial solution and if it takes the value zero, the original system is represented.

Along with ensuring convergence, Tx stepping avoids the undesirable low voltage solutions for the steady-state solution of the combined T&D network by pushing the solution toward high-voltage meaningful result. Based on the manner in which Tx stepping "shorts" the system, the initial solution corresponds to high system voltages, and each subsequent step of the homotopy approach continues and deviates slightly from this initial solution, thereby ensuring convergence to the high voltage solution for the original problem.

## IV. COMBINED T&D SIMULATION SETUP

### A. General Framework

In equivalent circuit framework for combined T&D analysis, we first model the transmission and distribution grid elements using currents and voltages state variables, as illustrated in case of PQ load model in Section II. After, we combine the resulting transmission positive sequence equivalent circuits with distribution three-phase circuits using a coupling port that is discussed below. Then, we apply NR-based methods to solve the set of non-linear equations defined by the aggregated T&D circuit either in parallel on multiple cores (on large machines or in the cloud) or as one large problem on a single core.

### B. Coupling Port for Positive Sequence and Three-Phase Network

For steady-state analysis of combined T&D networks, coupling of positive sequence transmission networks and three-phase distribution networks is first required. We develop a "coupling port" circuit to achieve this.

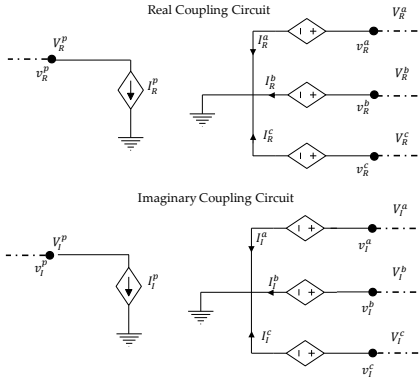

*Figure 3: Coupling port for joint transmission and distribution analysis.*

Fig. 3 graphically depicts the equivalent circuit for the coupling port. The set of nodes $(v^p \subseteq \mathcal{V} = \{v_R^p, v_I^p\})$ on left of the figure represent the nodes for the coupling circuit on the transmission side, whereas the set of nodes on the right of the figure $(v^{abc} \subseteq \mathcal{V} = \{v_R^a, v_I^a, v_R^b, v_I^b, v_R^c, v_I^c\})$ represent the nodes on the distribution end. The controlled current sources ($I_R^p$ and $I_I^p$) and controlled voltage sources ($V_R^a$, $V_R^b$, $V_R^c$, $V_I^a$, $V_I^b$ and $V_I^c$) represent the coupling variables between the transmission and distribution sub-circuits. For instance, the controlled current sources ($I_R^p$ and $I_I^p$) represent the positive sequence currents consumed by the distribution sub-circuit, whereas controlled voltage sources represent the nodal voltages for the distribution grid ($v^{abc}$) that mirrors the transmission voltages at the POI. Importantly, this port enables connection of any number of transmission and distribution networks without loss of generality and together they include the set of nodes $\mathcal{V}^p$ and $\mathcal{V}^{abc}$ to represent all nodes for coupling circuits at transmission and distribution side, respectively.

The theory of symmetrical components [26] is used to derive the expressions for coupling currents and voltages required to model the port. The positive sequence model of the power grid assumes balanced operation of the grid, and therefore, ignores the zero and negative sequence components of voltages and currents. Hence, by ignoring the zero and negative sequence currents consumed by the distribution feeder at POI, we first calculate the positive sequence currents consumed by the distribution feeders and then represent the same using coupled sources at the POI on transmission grid.

$$\begin{bmatrix} I_R^z \\ I_I^z \\ I_R^p \\ I_I^p \\ I_R^n \\ I_I^n \end{bmatrix} = \begin{bmatrix} 1 & 0 & 1 & 0 & 1 & 0 \\ 0 & 1 & 0 & 1 & 0 & 1 \\ 1 & 0 & \alpha^2 & 0 & \alpha & 0 \\ 0 & 1 & 0 & \alpha^2 & 0 & \alpha \\ 1 & 0 & \alpha & 0 & \alpha^2 & 0 \\ 0 & 1 & 0 & \alpha & 0 & \alpha^2 \end{bmatrix}^{-1} \begin{bmatrix} I_R^a \\ I_I^a \\ I_R^b \\ I_I^b \\ I_R^c \\ I_I^c \end{bmatrix} \quad (9)$$

Similarly, the distribution end voltages as a function of transmission POI voltages are calculated via:

$$\begin{bmatrix} V_R^a \\ V_I^a \\ V_R^b \\ V_I^b \\ V_R^c \\ V_I^c \end{bmatrix} = \begin{bmatrix} 1 & 0 & 1 & 0 & 1 & 0 \\ 0 & 1 & 0 & 1 & 0 & 1 \\ 1 & 0 & \alpha^2 & 0 & \alpha & 0 \\ 0 & 1 & 0 & \alpha^2 & 0 & \alpha \\ 1 & 0 & \alpha & 0 & \alpha^2 & 0 \\ 0 & 1 & 0 & \alpha & 0 & \alpha^2 \end{bmatrix} \begin{bmatrix} 0 \\ 0 \\ V_R^p \\ V_I^p \\ 0 \\ 0 \end{bmatrix} \quad (10)$$

Importantly, if unbalanced operation is expected at the high voltage transmission system level, then one must construct the three-phase equivalent circuit of the transmission system and couple it directly with the three-phase equivalent circuit of the distribution system at the POI. This can be done via an equivalent circuit approach by following the formulation set forth in [15]-[17]. However, the analysis of an unbalanced three-phase transmission network is beyond the scope for this paper.

## V. SIMULATION ALGORITHM

In the previous sections we demonstrated how the positive-sequence transmission and the three-phase distribution networks are first created and then coupled via the use of a proposed coupling port. In this section we explore different approaches for solving the coupled aggregated circuits. A simple approach is to solve the coupled T&D network's non-linear equations on a single machine using NR direct solve with the help of circuit simulation techniques. This approach works well until the size of combined T&D network becomes too large, thereby necessitating the use of parallel computing approaches. Therefore, in this paper we study two approaches for obtaining the steady-state solution of combined T&D networks:

i. Combined T&D simulation on a single machine using NR direct solve.
ii. Combined T&D simulation on distributed cores using Gauss-Seidel Newton (GSN) method.

### A. Combined T&D simulation on a single machine

Fig. 4 details the step-by-step algorithm for obtaining the steady-state solution of the combined T&D network using NR direct solve on a single core. The direct NR solver begins by parsing the input data for the transmission as well as distribution networks. It proceeds to fill (i.e. stamps) the parameters of the linear and linearized non-linear models from the parsed data into the system matrix. The coupling circuits are then stamped in the system matrix to aggregate the T&D networks. The linearized system matrix is then solved via NR method with iterative update of the non-linear stamps until a solution is reached. During NR iterations, circuit simulation parameters and any corresponding homotopy circuits are also stamped as needed to achieve convergence. To achieve robust yet fast convergence, in our solver from the practical point of view, the available initial conditions in the input file are first used as the initial conditions and homotopy factor λ is set to 0 (representing the original problem). However, in cases where the system is ill-conditioned or lacking a good initial guess, if the error profile is observed to be diverging or non-converging (based on developed heuristics), then the solver begins to gradually increase the homotopy factor λ until a solution is found. Once the solution to the trivial problem is obtained, the homotopy factor is dynamically scaled back down to 0 based on developed heuristics. Finally, any control variables are adjusted as needed and the NR loops are re-run until convergence.

*Figure 4: Combined T&D simulation with direct NR solve.*

**ALGORITHM**
1: **procedure:**
2: **parse** transmission and distribution networks
3: **couple** the networks at POI with coupling ports
4: **stamp** positive sequence and three-phase models of the coupled system
5: **stamp** the coupling port circuit models
6: **while** not converged **do** NR step with limiting**:**
7:    **update** non-linear stamps
8:    **check** error profile
9:    **if** error profile diverging**:**
10:      **adjust** homotopy factor (λ)
11:    **end if**
12:    **if** change in λ: **update** homotopy stamps
13: **end while**
14: **check** control variables, **if** any violated: **update** and **goto** step 5

### B. Combined T&D Simulation using Distributed Framework

When solving large sized coupled T&D systems with hundreds of distribution feeders connected to one or more transmission networks, the computational capacity and the system memory of a single machine may be insufficient. Beyond a certain sized coupled system, the combined T&D simulation becomes computationally impractical on a single machine due to the large size of the solution matrix that requires larger memory and more compute power. Therefore, to address this limitation, we use a parallel simulation framework with the use of distributed cores that are available either on large servers or in a computing cloud.

*1) Domain based decomposition of T&D network*

In circuit simulation, system matrices of large integrated circuit type problems are representable in a special bordered block diagonal form (BBDF), which has led to significant research toward solving such problems in parallel [28]-[33]. The solution matrix for the combined T&D simulation problem is inherently in the same BBD form, which exists due to the hierarchical nature of coupling between the various T&D networks. Therefore, the developed theory for parallel simulation in circuit simulation field can also be directly applied to this problem.

To begin solving a large combined T&D simulation problem in parallel, we first decompose the combined T&D network into smaller networks. We use domain-based decomposition [27]-[28] to achieve this, then represent the decomposed network in BBD form. Finally, we apply a Gauss-Seidel-Newton (GSN) algorithm [30], [31] to solve the torn BBDF-representable combined T&D problem robustly. In circuit simulation field, this method has been shown to have guaranteed convergence [32] for circuits that exhibit certain properties and structure. We investigate and define these properties for the power grid circuits.

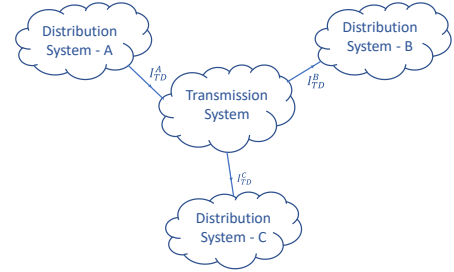

*Figure 5: Weakly coupled transmission and distribution network.*

Successful application of domain-based decomposition requires that the sub-systems with the aggregated system are weakly coupled. There is a natural weak coupling between the different transmission and distribution networks in the electric grid, as shown for a simple example in Fig. 5. This coupled network can be divided into a set of sub-systems ($S$) by the branch tearing technique at the coupling points (represented by nodes and edges of the coupling circuit) between the transmission and distribution network, as shown in Fig. 6. To generalize this and demonstrate mathematically, consider the coupled T&D network with the following function form:

$$\mathcal{F}(V_R, V_I) = 0 \qquad (11)$$

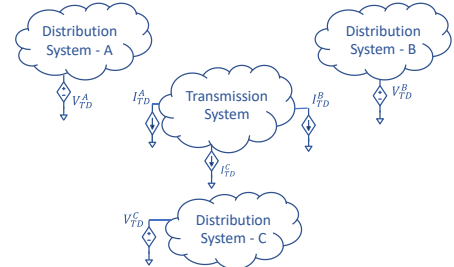

*Figure 6: "Torn" transmission and distribution sub-systems.*

The large coupled T&D network given by (11) is torn into $m$ independent sub-circuits that consist of the internal variables ($x^{int}: \{V_R^{int}, V_I^{int}\}$) that are only function of circuit elements within its own sub-circuit and the external variables ($x^{ext}: \{V_R^{ext}, V_I^{ext}\}$) that are functions of circuit elements in the other sub-circuits [28]. The decomposed sub-circuits then have the following function form:

$$\mathcal{F}_{int}(V_R^{int}, V_I^{int}, V_R^{ext}, V_I^{ext}) = 0 \qquad (12)$$
$$\mathcal{F}_{ext}(V_R^1, V_I^1, \dots, V_R^m, V_I^m, V_R^{ext}, V_I^{ext}) = 0 \qquad (13)$$

for $int = 1, \dots, m$, where $\{V_R^{int}, V_I^{int}\} \in \mathbb{R}^{n_i}$ are the internal nodal voltages of sub-circuits, and $\{V_R^{ext}, V_I^{ext}\} \in \mathbb{R}^{n_e}$ are the external nodal voltages. Importantly, the set of sub-circuits are only considered weakly coupled if for each sub-circuit $S \in \mathbf{S}$:

$$\dim(x_S^{ext}) \ll \dim(x_S^{int}) \qquad (14)$$

Next, we represent the system matrix $J$ of the torn network in BBDF form, as shown in Fig. 7, and apply GSN to solve it. In Fig. 7, the block diagonal terms in the BBDF structured system matrix $(T, D_A, D_B, D_C)$ represent the elements for the decomposed sub-circuits $(S = \{\mathcal{F}_T, \mathcal{F}_{D_A}, \mathcal{F}_{D_B}, \mathcal{F}_{D_C}\})$ that are functions of sub-circuit's internal parameters $(x_S^{int})$, whereas the off-diagonal terms in the vertical right of the matrix i.e. $(tt', td_a, td_b, td_c)$ are system elements that are functions of sub-circuit's circuit external variables $(x_S^{ext})$. Remaining elements in the bottom of the matrix $(TT', TD_a, TD_b, TD_c)$ map the behavior of the coupling circuits. For simulations of combined T&D networks, the systems equations are formulated based on modified nodal analysis (MNA). For MNA constructed BBDF matrix, the block diagonal elements consist of terms for each T&D sub-system model excluding those for coupling circuits. The bottom elements include the current and voltage functions of the coupling circuits whereas the right band of the matrix add current terms flowing out of the coupling circuit nodes to satisfy KCL at those nodes.

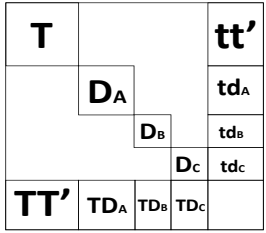

Figure 7: Bordered Block Diagonal structure for joint transmission and distribution system.

*2) Gauss-Seidel-Newton Approach*

We apply Gauss-Seidel-Newton (GSN) method [30], [31] to solve the set of sub-systems given by decomposed T&D sub-circuits. The subsystems are chosen such that the number of internal nodes for each sub system $(x_S^{int})$ are far more than the number of external coupling nodes $(x_S^{ext})$. The GSN algorithm is a two-step algorithm. Within the inner loop, the set of independent sub-systems $(S)$ are solved in parallel using the block NR algorithm until convergence or for a limited number of iterations. In this inner loop, the external coupling variables $(x^{ext})$ are kept constant for each sub-circuit, whereas the internal variables $(x^{int})$ are solved for iteratively. In the outer loop, the external coupling variables from each sub-system are distributed to other sub-systems via a Gauss step, and the inner loop of NRs are performed again. This iterative algorithm is then repeated until the error difference of external coupling variables communicated between the consecutive outer loops (epochs) are within a certain tolerance.

Figure 8: Combined T&D simulation with GSN implemented in parallel.

**ALGORITHM**

1: **procedure:**
2: **parse** combined T&D network into individual sub-circuits $S_i$
3: **initialize** external variables $x^{ext}$ for each $S_i$
4: **while** $\Delta x^{ext} < tol$ **do** Gauss step, for $\forall S_i \in S$ in parallel:
5: **stamp** linear and non-linear models within the sub-circuit $S_i$
6: **initialize** $x_i^{int}$ and assign and keep constant $x_i^{ext}$
7: **while** not converged **do** NR step with limiting:
8: **update** $x_i^{int}$ corresponding to any non-linear stamps
9: **check** error profile
10: **if** error profile diverging:
11: **adjust** homotopy factor $(\lambda)$
12: **end if**
13: **if** change in $\lambda$: **update** homotopy stamps
14: **end while**
15: **check** control variables, **if** any violated: **update** and **goto** step 6
16: **update** external variables $(x^{ext})$ and goto step 4
17: **end while**

While applying GSN to combined T&D simulation problem, solving of individual sub-systems within the inner loop using NR equates to running independent instances of power flow and three-phase power flow in parallel. A necessary condition for convergence of the GSN algorithm is that each of these independent power flows and three-phase power flows for the individual sub-systems that are solved at every Gauss step produce a correct meaningful solution; absence of which could result in divergence or breakdown of the complete algorithm. We have previously demonstrated in [17] and [35] that use of circuit simulation methods along with other adapted circuit theoretic methods can ensure convergence for any feasible or infeasible power flow or three-phase power flow circuit, which makes these methods well suited for the proposed GSN algorithm. The adjustment of the corresponding circuit simulation parameters in this algorithm is further detailed in Section V for direct NR algorithm and is directly applicable to parallel GSN algorithm. Furthermore, this approach avoids numerical instability issues that may arise due to wide spread of parameters between the T&D grid elements by solving the T&D elements separately in different sub-systems. Fig. 8 details the implementation of GSN algorithm for combined T&D algorithm.

## VI. Convergence Properties of GSN Approach

Suppose the solution matrix of a large joint interconnected transmission and distribution network is given by:
$$JV = I \quad (15)$$
where solution matrix $J$ has the BBD form given in Fig. 7. To further explore the convergence properties of GSN algorithm on this form of solution matrix, the matrix $J$ can be split into two components that are given by:
$$J = M - N \quad (16)$$

In general, for the Gauss-Seidel-Newton (GSN) algorithm to guarantee convergence for the decomposed matrix $J$, the spectral radius of the iteration matrix $(\rho(M^{-1}N))$ needs to be less than 1. However, a less strict condition that requires the solution matrix $J$ to be point-wise strictly diagonal dominant is often sufficient; i.e.
$$\sum_{i \neq j}^{n} |a_{ij}| \leq |a_{ii}|, \text{ for all } i \quad (17)$$
where $a_{ij}$ is a value in the $J$ matrix for $i^{th}$ row and $j^{th}$ column. For general circuit-based problems, satisfying these conditions can be hard, therefore we explore less restrictive *topological* conditions for GSN algorithm that are applicable to this problem due to the special structure of the solution matrix $J$.

Authors in [32] develop milder *topological* conditions for guaranteed convergence of GSN algorithm when simulating block relaxation circuit problems with special BBDF matrix structure under the hold of certain assumptions. The structure of the solution matrix for combined T&D steady-state simulation mimics that of BBDF solution matrices used within block relaxation circuit problems in [32] and hence same set of less restrictive *topological* conditions for convergence apply to this problem. To further discuss these milder conditions, we define a set of feedback $\mathcal{V}_{fb} \subseteq \mathcal{V}$ and feedforward $\mathcal{V}_{ff} \subseteq \mathcal{V}$ nodes within the combined T&D circuit in Appendix A. These nodes have special properties in terms of convergence of GSN method and refer to the set of nodes within the coupling circuits of the aggregated circuit as shown in Appendix A.

For an equivalent T&D circuit whose solution matrix $J$ has the BBD structure shown in Fig. 7, existence of a capacitance with a large enough value at a small sub-set of system nodes that are given by the feedback nodes ($\mathcal{V}_{fb} \subseteq \mathcal{V}$) can guarantee convergence for the aggregated system via GSN algorithm. In our approach, the aggregated equivalent circuit is torn into multiple sub-circuits at coupling ports that consist of following nodes $\{\mathcal{V}^{abc}, \mathcal{V}^p\}$. Feedback $\mathcal{V}_{fb}$ nodes are shown in Appendix A to be equivalent to the set of coupling circuit's nodes on the distribution side $\mathcal{V}^{abc}$ and feedforward nodes $\mathcal{V}_{ff}$ are shown to be equivalent to coupling circuit's nodes on the transmission side $\mathcal{V}^p$. Therefore, if the aggregated circuit is torn such that the distribution nodes $v^{abc}$ of the coupling circuit all have capacitors of large enough value then the GSN algorithm can be shown to be provably convergent. This is a much less restrictive *topological* sufficient condition for convergence when compared against the strict diagonal dominance condition for a general solution matrix that requires a large valued capacitor from each node in the system to ground.

As an alternative approach, if the system matrix $J$ can be represented as a non-symmetric positive definite matrix ($J \succ 0$) as is the case with many power networks [36]-[37] and has the BBD structure as is the case in our combined T&D approach, then a method presented in [33] can be applied to ensure convergence. [33] ensures convergence for the power flow network-based problems via GSN by partitioning the BBDF matrix $J$ such that the spectral radius of the iteration matrix corresponding to the partitioned system is ensured to be less than one. The work in [33] partitions the solution matrix $J \in \mathbb{R}^{NxN}$ into $M \in \mathbb{R}^{NxN}$ and $N \in \mathbb{R}^{NxN}$ such that $J = M - N$, where $M$ is a block diagonal matrix capturing the interactions between the internal variables of each block sub-circuit and $N$ is the off-diagonal matrix that captures the communication between the variables of other sub-circuits. To ensure convergence of GSN, the method introduces a diagonal matrix $\bar{E} \in \mathbb{R}^{NxN}$ with $i^{th}$ entry of the matrix equal to:

$$\bar{E}_{ii} = \alpha \sum_{j=1}^{N} E_{ij} \quad (18)$$

such that the matrices M and N are modified as follows:
$$M = D + \alpha \bar{E} \quad (19)$$
$$N = \alpha \bar{E} - E \quad (20)$$

where, $J = D + E$. It is shown in [33] that by choosing the value of $\alpha = \frac{1}{2}$, the algorithm can ensure convergence for the partitioned system. To expand further, [33] shows that to prove that the spectral radius $\rho(M^{-1}N) < 1$, it is sufficient to have $J = M - N \succ 0$ and $M + N \succ 0$. $M - N \succ 0$ is positive-definite based on the initial assumption and $M + N$ can be shown to be positive definite by showing that $D + 2\alpha \bar{E} + E$ is strictly diagonal dominant.

Furthermore, it should be noted that GSN is not the only available algorithm for solving BBD structured circuit matrices in parallel. For instance, direct NR method can solve BBD structured circuit problems in parallel by applying distributed Schur's complement [38] to extract the exact solution of the linearized $J_l V = I_L$ in parallel at each step of NR.

## VII. RESULTS

The NR and GSN solvers were implemented in our prototype tool SUGAR (Simulation with Unified Grid Analyses and Renewables) to perform the simulations on the combined T&D networks. In the following results, the test networks for the transmission systems include the 9241 node PEGASE test system [39], ACTIVSg70k and ACTIVSg10k synthetic US networks [40], and the Eastern Interconnection of the US electric grid. The distribution networks include 8k+ nodes publicly available taxonomy feeder test cases [34] and 145 node three-phase model [17].

### A. Robustness due to Circuit Simulation Methods

Robust convergence of combined T&D circuits in distributed approach is dependent on robust convergence of individual transmission and distribution sub-circuits. Therefore, to demonstrate the efficacy our approach, we first simulate "hard-to-solve" transmission and distribution test cases in our approach with and without the use of circuit simulation methods. We simulate the same test case in commercial tools as well. We use a publicly available ACTIVSg70k.raw network [40] as the test network for the transmission grid. We modify the network by scaling it's loading factor (LF) to 0.8 (gen. real power and loads real and reactive power are scaled per loading factor). We then solve the network for three scenarios (i) in SUGAR w/o circuit simulation methods, (ii) in SUGAR with circuit simulation methods and (iii) in commercial tool. We document the results in Table 1 wherein it is shown that we solve the positive sequence network only with the use of circuit simulation methods and are unable to solve this system without the use of circuit simulation methods or in a commercial tool. We solve a similar "hard-to-solve" 145 node three-phase distribution test case as well from [17] with loading factor of 1.01. As was the case with transmission network, we can solve this system with the use of circuit simulation methods only.

TABLE 1: SUGAR WITH AND W/O CIRCUIT SIMULATION METHODS

| Case Name | Network Type | SUGAR w/o simulation methods | SUGAR with simulation methods | Commercial Tool |
|---|---|---|---|---|
| ACTIVSg70k @ 0.8 LF | Positive Seq. | Diverge | Converge | Diverge |
| Case145@ 1.01 LF [17] | Three-phase | Diverge | Converge | DNR* |

*Unable to run the system in GridLab-D due to non-negative impedance

These results demonstrate that in order to solve "hard-to-solve" transmission and distribution test cases from initial conditions far from solution, circuit simulation methods were essential in our approach whereas the commercial tools were unable to solve them at all. During combined T&D simulation using GSN if any of underlying sub-circuits fails to converge then the whole framework would fail. The ability of our approach to robustly solve individual "hard-to-solve" sub-circuits validates the robustness of our approach over other co-simulation frameworks.

## B. Combined T&D Simulation on a Single Machine

Here we discuss results for steady-state simulations on combined T&D networks using direct NR solve on a single machine.

### 1) Experiment 1

In the first experiment, the transmission network is represented by a 9241 node PEGASE test system that is coupled to a distribution grid modeled by a taxonomical feeder test case (R5-35.00-1) at the point of interconnection (POI) given by transmission system's node 2159. In this experiment, the original distribution feeder model is modified to include distributed energy resources (DERs) in roughly 20% of the system nodes that contain electrical loads. The net capacity of DERs at each node is kept variable and is modified throughout the experiment.

The goal of this experiment is two-fold: i) to demonstrate that higher capacity of distribution loads can be supplied with higher penetration of DERs and ii) to demonstrate that more resilient grid voltages can be obtained by higher penetration of DERs during both normal and contingency operation.

To obtain the base (w/o DERs) maximum loading for the combined T&D system studied within this experiment, we first develop the PV curve for the node voltages at the POI by varying the loading factor (where loading factor of value 1 denotes normal loading) of the distribution feeder, as shown in Fig. 9 (left). We repeat this analysis on the coupled system with a loss of a generator on the transmission grid that is in close vicinity of the POI. As seen in the Fig. 9 (left side), for the base case with no DERs, the voltages after the contingency has occurred have dipped below 0.75 pu for any loading factor greater than 0.38x, resulting in increased likelihood of voltage collapse.

To supply the full load in the distribution feeder such that the POI voltages remain above 0.75 pu for all loading factors up to 1.3x, we increase the penetration of DERs in the distribution system. We simulate the contingency and normal cases again and show the results in Fig. 9 (right side). With the availability of DERs in the system, the voltages are above 0.75 pu under normal as well as contingency scenarios for all loading factors up to 1.3x, while being able to supply greater than rated load of the distribution feeder without system collapse.

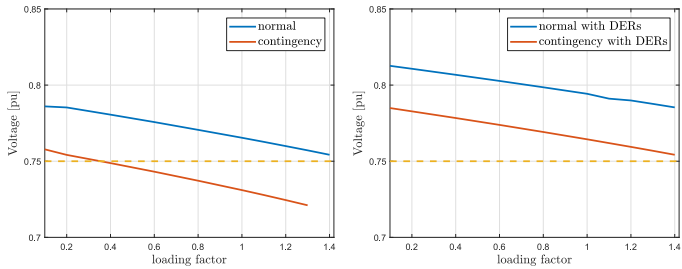

*Figure 9: Combined T&D simulation for a 9241-nodes transmission network coupled to 8000+ nodes distribution network.*

### 2) Experiment 2

A similar experiment is performed on a larger more realistic test case. In this experiment, the 85k+ nodes Eastern Interconnection network of the U.S. transmission grid is modeled via a positive sequence network. The 8k+ nodes taxonomical three-phase test system is then coupled to a weak point on this transmission network where the voltages are highly sensitive to load currents. The primary goal of this experiment is to evaluate the minimum penetration of DERs required to supply the full load of the distributed feeder while ensuring that the sub-station voltage at the POI remains above 0.75 pu.

To first evaluate the maximum transfer capacity at the POI prior to voltage collapse, we gradually increase the loading factor of the distribution feeder until the system collapses. As seen in Fig. 10 (left), the system can only supply a fraction of the rated capacity (0.7x loading factor) prior to voltage collapse without any DERs in the distribution network.

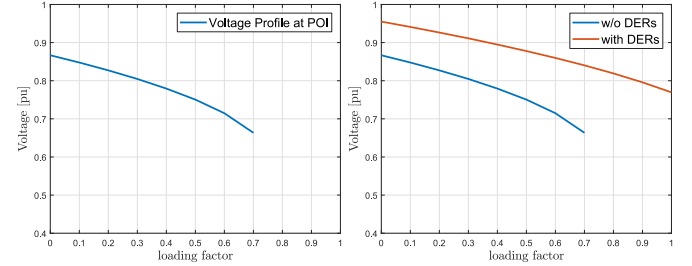

*Figure 10: Combined T&D simulation for a 9241-nodes transmission network coupled to 8000+ nodes distribution network.*

As a remedial action, the penetration of DERs in the system is increased until the transmission grid can supply the full load of the distribution feeder while keeping the voltages at the POI above 0.75 pu. As in the case of the prior experiment, the DERs in the system are added to roughly 20% of the total system nodes that contain electric loads. A scaling factor is used to increase the penetration of DERs in the simulation. Fig. 10 (right) displays that with 20% penetration of distribution generation in the distribution grid, the transmission network can supply the full load while maintaining grid voltages above 0.75 at the interconnection sub-station.

## C. Combined T&D Simulation in Parallel

Here we describe steady-state simulations on combined T&D networks with the use of parallel GSN setup. For the next set of experiments, we use the following machine:

| Operating System | # Cores | Type of CPU |
|---|---|---|
| Red Hat Enterprise | 32 | Intel(R) Xeon(R) CPU E5-2620 v4 @ 2.10GHz |

### 1) Validation

To first validate the distributed parallel simulation framework (as described in second part of Section V), we compare the results obtained from the parallel framework using GSN against those produced by direct NR algorithm on a single core (as described in first part of Section V). To setup the comparison, we couple a ~8k node taxonomical distribution feeder [34] with a 9241 node PEGASE test case at the transmission node 2519. We then simulate the coupled system for varying loading factors of the distribution feeders for the following setups:

i. The coupled T&D network solved as a single matrix problem on a single core using direct NR algorithm.
ii. The coupled T&D network solved in parallel on multiple cores using GSN algorithm.

Fig. 11 shows that the results obtained from the single core NR setup compare well with those obtained from the parallel simulation setup using GSN thereby validating the distributed GSN approach. The difference between the two results is within the tolerance used for the GSN algorithm (1e-3).

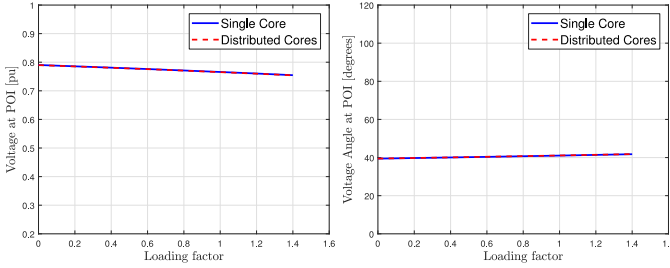

*Figure 11: Comparison of joint T&D simulation algorithms: i) Single machine setup using NR (in blue), ii) Parallel simulation on distributed cores using GSN (in red).*

*2) Combined T&D Simulations on Large Systems*

To perform the first large-scale experiment, 50 distribution feeders, each representing roughly 8k nodes, are coupled to a large realistic transmission network at different locations. The real Eastern Interconnection network of the US electric grid with roughly 85k+ nodes is used to represent the transmission network, and the taxonomical feeder test cases [34] are used to represent the set of distribution feeders. This problem represents a solution matrix size of roughly 3 million rank with a total of ~3x4,00,000 distribution nodes and ~85000 transmission nodes. We simulate for steady-state solution of this large network from flat start by applying GSN in parallel. In the end of the simulation, we document the grid sub-station voltages at POIs. The complete simulation took 11 Gauss steps wherein within each Gauss step, NR simulations run in parallel for each of the block sub-circuits took 5-10 iterations. In the final solution, the POI voltages were all found to be within the acceptable range of 0.8-1.2 pu and the complete simulation took less than a couple of minutes to converge with Tx-stepping method enabled. Fig. 12 shows the evolution of the sub-station voltages at the POI during the Gauss-step in the outer loop of the parallel combined T&D simulation.

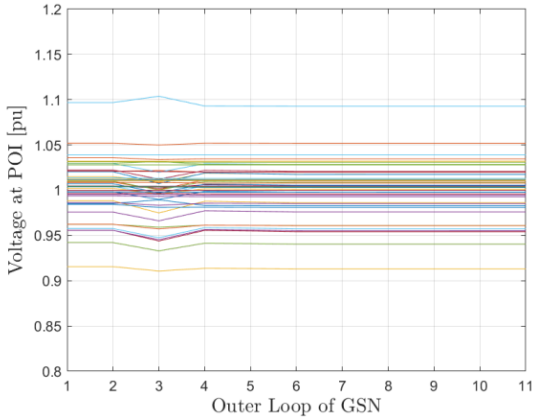

*Figure 12: Voltages at the POI in the outer loop of GSN.*

To show a similar result on a publicly available testcase, we couple a synthetic U.S. Eastern Interconnection sized transmission network ACTIVSg70k [40] with 100 taxonomical distribution feeders [34]. This combined T&D network has a solution matrix of size greater than 1.5 million and details for its construction are provided in Appendix B of this paper. To demonstrate convergence robustness of our approach on large combined T&D systems, we solved for the steady-state solution of this network when initialized from *flat-start*. The simulation took 448.29 second to run with total of 12 Gauss steps. The maximum and minimum voltage magnitude amongst the sub-station nodes where T&D systems interconnect is reported in Table 2 along with the node number at which it was observed.

TABLE 2: MAX. AND MIN. VOLTAGES FOR THE COMBINED T&D NETWORK

|  | Node Number | Magnitude |
|---|---|---|
| **Max. Voltage** | 24157 | 1.0420 |
| **Min. Voltage** | 27104 | 0.9651 |

*3) Scalability of GSN*

Herein, we demonstrate that the proposed GSN framework for distributed combined T&D simulations is highly scalable. To demonstrate this, we run an experiment wherein we construct five separate combined T&D systems, each representing an incrementally increasing system size. To achieve this, we couple a publicly available transmission grid network testcase ACTIVSg10k [40] with a set of increasing number of distribution feeders that are available in [34] and simulate the coupled T&D networks from *flat-start*. The number of distribution feeders connected to transmission network for the five combined T&D systems are in the set: {1, 10, 50, 100, 150} where the largest combined system has a solution matrix ($J$) size of greater than 2 million.

Fig. 13 shows the runtime of the GSN algorithm while performing combined T&D analysis against the dimension of the solution matrix in combined T&D network (on the left) and against the total number of distribution feeders that are coupled to the transmission network (on the right).

As is seen in the result the framework is highly scalable and robust. The framework can solve a test case with solution matrix ($J$) size greater than 2 million (wherein 150 distribution feeders are connected to the transmission system) in less than 5 minutes from flat-start.

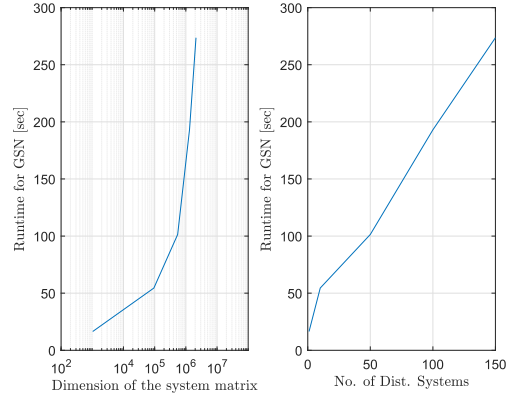

*Figure 13: Scalability of the GSN solver.*

## VIII. CONCLUSIONS

In this paper we proposed an equivalent circuit framework for obtaining the steady-state solution of the combined transmission and distribution networks. We demonstrated that the proposed framework can solve any coupled T&D network independent of the size or the complexity. We modeled the transmission grids with positive sequence network models and distribution grids with three-phase network models and further combined the two with proposed coupling circuits. We showed that the smaller-sized combined T&D networks (<1e6 nodes) can be solved on a single machine with direct NR solver whereas larger-sized combined T&D networks (>1e6 nodes) can be solved in parallel with parallel GSN solver on multiple cores available either on a large server or in the cloud.

## X. Appendix

### A. Appendix A

Here, we define and describe how the feedback nodes are identified for the aggregated circuit for the combined T&D network described within this paper. Per [32],

*Definition A-1*: A system variable $x_{kl}$ with $1 \leq k \leq \dim(s)$ and $1 \leq l \leq \dim(x_k)$ is called a feedback variable, such that for indices $i, j$, and $k < i \leq \dim(s), 1 \leq j \leq \dim(x_i)$, following condition for the term in non-linear system Jacobian $J$ is satisfied:

$$\frac{\partial J_{kl}}{\partial x_{ij}} \neq 0 \quad (A-1)$$

The node corresponding to the feedback variable is then called the feedback node $v_{fb}$.

*Definition A-2*: A system variable $x_{kl}$ with $1 \leq k \leq \dim(s)$ and $1 \leq l \leq \dim(x_k)$ is called a feedforward variable, such that for indices $i, j$, and $1 \leq i < k, 1 \leq j \leq \dim(x_i)$, following condition for the term in system Jacobian is satisfied:

$$\frac{\partial J_{kl}}{\partial x_{ij}} \neq 0 \quad (A-2)$$

The node corresponding to the feedforward variable is then called the feedforward node $v_{ff}$.

In our approach for combined T&D simulation, the coupling between transmission and distribution grids is always achieved via the coupling port that is described in Section IV-B and therefore the set of feedback ($\mathcal{V}_{fb}$) and feedforward ($\mathcal{V}_{ff}$) nodes within the combined T&D network are a sub-set of set of nodes $\{\mathcal{V}^p, \mathcal{V}^{abc}\} \subseteq \mathcal{V}$ within the coupling circuits. To further elaborate consider the coupling circuit in Fig. 3. If the sub-systems $S$ in the $J$ matrix are arranged such that transmission sub-system block is stamped in the top left of the matrix (i.e. $k = 0$, where $1 \leq k \leq \dim(s)$), then all the feedforward nodes ($\mathcal{V}_{ff}$) are equivalent to set of all nodes that include nodes of the coupling circuit on the transmission side ($\mathcal{V}_{ff} \equiv \mathcal{V}^p$) and all the feedback nodes $\mathcal{V}_{fb}$ are equivalent to the set of all nodes that include nodes of the coupling circuits on the distribution side ($\mathcal{V}_{fb} \equiv \mathcal{V}^{abc}$).

### B. Appendix B

Here we describe the construction of combined T&D network that is simulated in the Section VII.C.2 of this paper. To construct a combined T&D network of large dimension, we coupled a publicly available 70k node transmission test case ACTIVSg70k [40] with total of 100 distribution feeders that are available in [34]. The name of the distribution feeder and the transmission node to which the individual feeder is connected to in this network is given in the following table:

TABLE B-1: LARGE COMBINED T&D NETWORK DETAILS

| Dist. Case Name | Coupling Transmission Node |
|---|---|
| R1-12.47-3 | 43232, 27133, 43676, 27166, 43191, 24363, 60593, 23964, 27842, 24159, 24361, 27769, 27829, 26697 |
| R1-12.47-4 | 24260, 21055, 24037, 43355, 19324, 21081, 60262, 23938 |
| R2-12.47-1 | 43204, 43198, 24264, 24040, 43170, 32054, 20708, 19360 |
| R3-12.47-1 | 60577, 24057, 43186, 27120, 5983, 5940, 43762, 27168 |
| R3-12.47-2 | 27157, 17054, 27623, 60410, 24334, 43321, 43222, 27411 |
| R3-12.47-3 | 10642, 43415, 24124, 60291, 24002, 27174, 43188, 20325 |
| R4-12.47-1 | 27156, 43208, 24023, 43283, 43282, 43362, 24158, 43499 |
| R4-12.47-2 | 43193, 24157, 23997, 5929, 6232, 43324, 27666, 27103 |
| R4-25.00-1 | 24032, 24280, 24348, 19143, 27154, 26282, 60584, 24341 |
| R5-25.00-1 | 24120, 16768, 24267, 27104, 22558, 27822, 26700, 16156 |
| R5-35.00-1 | 24136, 24035, 24128, 24131, 43640, 31322, 24161, 27432, 43337, 16528, 26274, 43265, 37440, 27773 |

## XI. Biographies

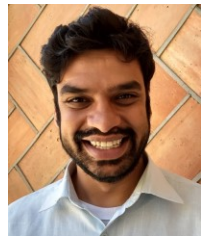

**Amritanshu Pandey** is a Senior Research Scientist at Pearl Street Technologies and an Adjunct Instructor at Carnegie Mellon University. He received his PhD. in Electrical and Computer Engineering from Carnegie Mellon University in 2019. Prior to joining as a doctoral student at Carnegie Mellon University, he worked as an electrical engineer at MPR Associates Inc. from 2012 to 2015. He has previously interned at Pearl Street Technologies, ISO New-England and GE Global Research. His research interests include modeling and simulation, optimization and control of power systems.

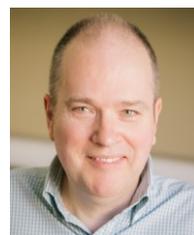

**Lawrence Pileggi** is the Tanoto professor of electrical and computer engineering at Carnegie Mellon University, and has previously held positions at Westinghouse Research and Development and the University of Texas at Austin. He received his Ph.D. in Electrical and Computer Engineering from Carnegie Mellon University in 1989. He has consulted for various semiconductor and EDA companies, and was co-founder of Fabbrix Inc., Extreme DA, and Pearl Street Technologies. His research interests include various aspects of digital and analog integrated circuit design, and simulation, optimization and modeling of electric power systems. He has received various awards, including Westinghouse corporation's highest engineering achievement award, a Presidential Young Investigator award from the National Science Foundation, Semiconductor Research Corporation (SRC) Technical Excellence Awards in 1991 and 1999, the FCRP inaugural Richard A. Newton GSRC Industrial Impact Award, the SRC Aristotle award in 2008, the 2010 IEEE Circuits and Systems Society Mac Van Valkenburg Award, the ACM/IEEE A. Richard Newton Technical Impact Award in Electronic Design Automation in 2011, the Carnegie Institute of Technology B.R. Teare Teaching Award for 2013, and the 2015 Semiconductor Industry Association (SIA) University Researcher Award. He is a co-author of "Electronic Circuit and System Simulation Methods," McGraw-Hill, 1995 and "IC Interconnect Analysis," Kluwer, 2002. He has published over 350 conference and journal papers and holds 40 U.S. patents. He is a fellow of IEEE.